\documentclass{iaus}
\usepackage{graphicx}

\title[Detection of planets by AO-assisted relative astrometry] 
{Detectability of planets in wide binaries by ground-based relative astrometry with AO}
\author[Neuh\"auser et al.]   
{Neuh\"auser R.$^1$ \thanks{rne@astro.uni-jena.de}, Seifahrt A.$^{1,2}$, R\"oll T.$^1$, 
Bedalov A.$^1$, Mugrauer M.$^1$}

\affiliation{$^{1}$Astrophysikalisches Institut, Universit\"at Jena, Schillerg\"a{\ss}chen 2-3, 07745 Jena, Germany\\
$^{2}$European Southern Observatory, Karl-Schwarzschild-Str. 2, 85748 Garching, Germany\\ [\affilskip]}

\pubyear{2007}
\volume{xxx}  
\pagerange{119--126}
\date{submitted October 2007}
\jname{Proceedings Title IAU Symposium}
\editors{W. Hartkopf, eds.}
\begin{document}

\maketitle

\begin{abstract}
Many planet candidates have been detected by radial velocity variations 
of the primary star; they are planet {\em candidates}, because of the 
unknown orbit inclination. Detection of the wobble in the two other dimensions,
to be measured by astrometry, would yield the inclination and, hence, 
true mass of the companions. We aim to show that planets can be confirmed or discovered
in a close visual stellar binary system by measuring the astrometric wobble
of the exoplanet host star
as periodic variation of the separation, even from the ground.
We test the feasibility with HD 19994, a visual binary with one radial
velocity planet candidate.
We use the adaptive optics camera NACO at the VLT
with its smallest pixel scale ($\sim 13$ mas)
for high precision astrometric measurements.
The separations measured in single 120 images taken within one night are shown to
follow white noise, so that the standard deviation can be devided by the square root
of the number of images to obtain the precision.
In this paper we present the first results and investigate the achievable precision in relative
astrometry with adaptive optics.
With a careful data reduction it is possible to achieve a relative astrometric precision
as low as 50 $\mu$as for a 0.6 arc sec, binary with VLT/NACO observations in one hour,
the best relative astrometric precision ever achieved with a single telescope from the ground.
The relative astrometric precision demonstrated here with AO at an 8-m mirror
is sufficient to detect the astrometric signal of the planet HD 19994 Ab
as periodic variation of the separation between HD 19994 A and B.
\end{abstract}

\firstsection 
\section{Introduction}
Since the radial velocity technique can yield only lower mass limits,
all planets (or planet candidates) found by this technique have to be
confirmed by other methods. Of $\sim 200$ radial velocity planet candidates
found so far, 14 have been confirmed by transit and two by astrometry,
see e.g. exoplanet.eu.
The two planets GJ 876 b and 55 Cancri d have been confirmed by astrometry
using the Hubble Space Telescope Fine Guidance Sensor 
(Benedict et al. 2002, McArthur et al. 2004),
with a precision down to 0.04 mas (milli arc sec)
sufficient to detect the wobble of the host star in the plane of the sky.

The astrometric displacement is given by
\begin{displaymath}
\theta\,[mas] = 0.960 \times \frac{a}{[5\,AU]} \times \frac{[10\,pc]}{d} \times
\frac{M_{pl}}{[M_{jup}]} \times \frac{[M_{\odot}]}{M_{\star}}
\end{displaymath}
with planet mass $M_{pl}$ in Jupiter masses at a semi-major axis $a$ (in units of 5 AU)
in a circular orbit around a host star with mass $M_{\star}$ in solar masses
at a distance $d$ (in units of 10 pc).

Here, we present both a feasibility study to determine the astrometric precision of
the Adaptive Optics (AO) camera NAos-COnica (NACO) of the ESO Very Large Telescope (VLT)
as well as first measurements in order to determine the mass of
HD 19994 Ab, a RV planet candidate in orbit around HD 19994:
The RV data suggest a lower mass limit of the
planet candidate of $m\cdot\sin{i} = 1.68~\mathrm{M_{Jup}}$
in an orbit with semi-major axis $a=1.42$ AU with an eccentricity of $e = 0.30 \pm 0.04$,
i.e. a $535.7 \pm 3.1$ day orbital period (Mayor et al. 2004). 
The full (peak-to-peak) astrometric wobble,
from Kepler's 3rd law, is at least 0.155 mas (for a circular orbit)
and 0.131 mas by taking the observed eccentricity
into account -- this is when assuming the minimum mass $m \cdot \sin i$ as true mass.

\section{Observations and data reduction}

We observed HD 19994 and a calibration binary (HD 19063) with the S13 camera (13.26 mas/pixel pixel scale,
14$^{\prime \prime}$ $\times$ 14$^{\prime \prime}$ FOV) in a NB\_2.17 narrow band filter. Short exposures of both 
target binaries where taken for one hour each, equally splitted up into two time slots
before and after meridian passage.
We obtained a total of 120 frames for HD 19994 and 117 frames for HD 19063.
Data reduction followed the standard technique of dark subtraction, division
by a flat field and the application of a bad pixel mask with the {\em eclipse} software.
In each single exposure (0.347 sec), we measured the separation between A and B
of both binaries. We have then made sure that the separations follow white noise
by a K-S test (see a future paper by Neuh\"auser et al. for details). Hence, we can
devide the standard deviation of the mean by the square root of the number of
measurements to obtain the precision of the separation measurement,
the {\em precision}, not the {\em accuracy}. For the latter, we 
would need to know the true pixel scale. See figure 1 and table 1.

\begin{figure}
\centering
\includegraphics[width=12.3cm]{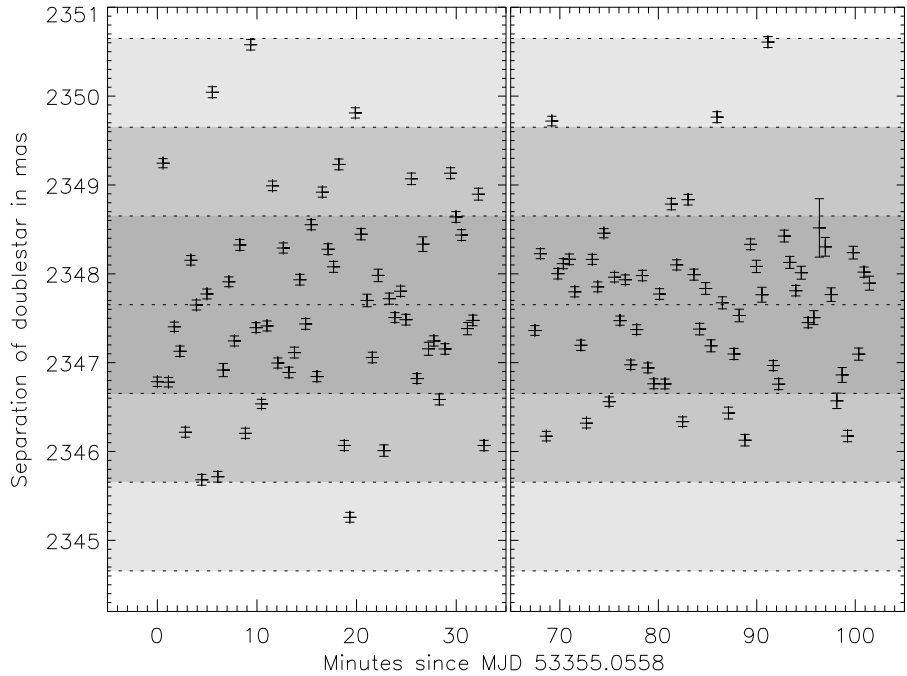}
\caption{Measured separation for HD 19994 (similar for HD 19063).
Error bars denote 1$\sigma$ error for individual measurements. 
The standard deviation of the complete dataset is shown as
dashed lines and grey-shaded areas
for 1, 2, and 3$\sigma$, respectively.
The standard deviation of the mean, thus the precision of the
measured binary separation, is $0.998/\sqrt{120}~mas~=~91.2~\mu as$.}
\end{figure}

\begin{table}[hb!]
\caption{Astrometric results from the measurements
of the binaries HD 19994 and HD 19063. K-S test values
higher than 0.6 are interpreted as
indicating white noise.}
\label{data}
\centering
\begin{tabular}{c c c c c r}
\hline
Target &Mean sepa-&Standard&N&K-S&Precision \\
&ration (mas)&deviation&   &test&achieved \\
\hline
HD 19994& 2347.652 & 0.998 mas& 120 & 0.69 &  91.2 $\mathrm{\mu as}$\\
        & 2347.632 & 1.110 mas& 60  & 0.85 & 143.3 $\mathrm{\mu as}$\\
        & 2347.673 & 0.882 mas& 60  & 0.64 & 113.9 $\mathrm{\mu as}$\\
HD 19063&          &       &     &      &       \\
        & 649.628  & 0.377 mas& 57  & 0.74 &  49.9 $\mathrm{\mu as}$\\
       & 649.596   & 0.391 mas& 60  & 0.48 & (failed) \\
\hline
\end{tabular}
\end{table}

Differential chromatic refraction (DRC) or Allan noise 
(see Pravdo \& Shaklan 1996) do not matter here,
because we observe in the near IR with a narrow-band filter (hence no DCR)
and with AO (hence no Allan noise).

\section{Results and conclusions}

We have observed two bright visual binaries (0.6 to 2.4 arc sec separation)
with VLT/NACOs smallest pixel scale (13.26 mas/pixel)
by taking $\sim 120$ short (0.347 s) exposures per binary within a few hours,
separated in four bins of $\sim 60$ exposures each
(15 December 2004, ESO program 075.C-0288.A).
We could confirm that the separations measured follow white noise
(in three out of four bins), so that we are allowed to divide the
mean of the separations by the square root of the number of
measurements to obtain the precision of the separation measurement.
By doing so, we measured the separation between HD 19994 A and B
to be $2347.652$ mas (milli arc seconds) with the relative precision
of $91.2 \mu as$ (micro arc seconds).
For HD 19063, we obtained $649.628$ mas
with a precision being as low as $49.9 \mu as$.
These are the most precise measurements in relative astrometry
obtained with single-aperture telescopes from the ground,
sufficient to detect the astrometric wobble due to a planet,
to be seen as periodic change in the binary separation.

Such a high precision as shown here can be applied not only to
measurements of masses of previously detected RV planets (or candidates),
but also to search for new planets in wide binaries, or even single stars with
a bright (back- or foreground) star within the isoplanatic angle.
Other possible applications are orbit determinations in stellar multiples,
ground-based parallaxe and proper motion measurements,
observations of expanding, contracting, or rotating star clusters,
etc.

The main limitation of this technique is the need for a bright
star near the target, namely within the isoplanatic angle.
And, as long as no pixel scale calibrator is available with a precision
down to 1/100000, the measurement can achieve this high precision
only for relative astrometry, but not in absolute terms.

\end{document}